\documentclass[twocolumn]{aastex631}
\usepackage{amsmath,mathtools,mathrsfs,cleveref, graphicx,float}
\usepackage[caption=false]{subfig}
\usepackage[T1]{fontenc}
\usepackage{makecell}
\usepackage{multirow}
\usepackage{hyperref}
\usepackage{color,soul}
\usepackage{nicefrac}

\hypersetup{
  colorlinks=true,        
  linkcolor=blue,         
  citecolor=magenta
}

\definecolor{forestgreen}{RGB}{34,180,34}

\begin{document}
\title{{\tt Mahakala}: a {\tt Python}-based Modular Ray-tracing and Radiative Transfer Algorithm for Curved Space-times}
\author[0009-0005-0398-3748]{Aniket Sharma}
\affiliation{Indian Institute of Science Education and Research Mohali. Sector 81, SAS Nagar, Mohali, PO Manauli, Punjab 140306, India}
\author[0000-0003-2342-6728]{Lia~Medeiros}
\affiliation{Center for Gravitation, Cosmology and Astrophysics, Department of Physics, University of Wisconsin–Milwaukee, P.O. Box 413, Milwaukee, WI 53201, USA}
\affiliation{Department of Astrophysical Sciences, Peyton Hall, Princeton University, Princeton, NJ, 08544, USA}
\affiliation{School of Natural Sciences, Institute for Advanced Study, 1 Einstein Drive, Princeton, NJ 08540, USA}
\author[0000-0001-6952-2147]{George~N.~Wong}
\affiliation{School of Natural Sciences, Institute for Advanced Study, 1 Einstein Drive, Princeton, NJ 08540, USA}
\affiliation{Princeton Gravity Initiative, Princeton University, Princeton, New Jersey 08544, USA}
\author[0000-0001-6337-6126]{Chi-kwan~Chan}
\affiliation{Steward Observatory and Department of Astronomy, University of Arizona, 933 N. Cherry Ave., Tucson, AZ 85721}
\affiliation{Data Science Institute, University of Arizona, 1230 N. Cherry Ave., Tucson, AZ 85721}
\affiliation{Program in Applied Mathematics, University of Arizona, 617 N. Santa Rita, Tucson, AZ 85721}
\author[0000-0002-7232-101X]{Goni~Halevi}
\affiliation{Department of Physics, Illinois Institute of Technology, Chicago, IL 60616, USA}
\affiliation{Center for Interdisciplinary Exploration and Research in Astrophysics (CIERA), Northwestern University, Evanston, IL 60201, USA}
\author[0000-0003-2131-4634]{Patrick~D.~Mullen}
\affiliation{CCS-2, Los Alamos National Laboratory, Los Alamos, NM}
\affiliation{Center for Theoretical Astrophysics, Los Alamos National Laboratory, Los Alamos, NM}
\author[0000-0001-5603-1832]{James~M.~Stone}
\affiliation{School of Natural Sciences, Institute for Advanced Study, 1 Einstein Drive, Princeton, NJ 08540, USA}
\correspondingauthor{Lia Medeiros}\email{lia2@uwm.edu}

\begin{abstract}
We introduce {\tt Mahakala}, a {\tt Python}-based, modular, radiative ray-tracing code for curved space-times. We employ Google’s {\tt JAX} framework for accelerated automatic differentiation, which can efficiently compute Christoffel symbols directly from the metric, allowing the user to easily and quickly simulate photon trajectories through non-Kerr spacetimes. {\tt JAX} also enables {\tt Mahakala} to run in parallel on both CPUs and GPUs. {\tt Mahakala} natively uses the Cartesian Kerr-Schild coordinate system, which avoids numerical issues caused by the pole in spherical coordinate systems.
We demonstrate {\tt Mahakala}'s capabilities by simulating 1.3~mm wavelength images (the wavelength of Event Horizon Telescope observations) of general relativistic magnetohydrodynamic simulations of low-accretion rate supermassive black holes. The modular nature of \texttt{Mahakala} allows us to quantitatively explore how different regions of the flow influence different image features. We show that most of the emission seen in 1.3~mm images originates close to the black hole and peaks near the photon orbit. We also quantify the relative contribution of the disk, forward jet, and counter jet to 1.3~mm images.
\end{abstract}
 
\section{Introduction}\label{sec:intro}

Accurate integration of null geodesics through curved spacetimes is crucial for modeling the observable electro-magnetic (EM) signature of accreting black holes. Ray-tracing through the curved spacetimes near black holes dates back to \citet{Bardeen_single_author}, \citet{Cunninggham_single_author}, and \citet{Luminet}, where the authors calculated the images of these objects for the first time. Comparing theoretical black hole accretion models with observations additionally requires solving the radiative transfer equation along photon trajectories.

With the advent of general relativistic magnetohydrodynamic (GRMHD) simulations of accreting black holes (see e.g., \citealt{GRMHD_De_villiers, GRMHD_HARM, GRMHD_Noble, GRMHD_sadowski, GRMHD_Koral_2014, Athena++}), ray-tracing has become a standard tool for simulating accretion disk images and spectra via radiation post-processing. Radiative ray-tracing calculations have been used to study variability properties (see e.g., \citealt{Variability_1,Variability_2,Chan2015,Medeiros2017,Medeiros2018a,Medeiros2018b}), emission and absorption lines (see e.g., \citealt{Monte_Carlo_code}), spectra (see e.g., \citealt{Grmonty}), and radiative efficiency (see e.g., \citealt{Radiative_efficiency}). GRMHD simulations have also been indispensable in modeling and interpreting the recent high-resolution, horizon-scale observations of two low-luminosity supermassive black holes by the Event Horizon Telescope (\citealt{EHT_paper, EHT_2, EHT_3, EHT_4, M87_paper_5, EHT_paper_6, EHT_7, EHT_8, EHT_9, EHT_10, EHT_11, EHT_12, SgrA_paper_5, EHT_14, 2023ApJ...957L..20E, 2024A&A...681A..79E}). 

The choice of programming language used to develop these algorithms directly influence their performance and usability. While some codes (see, e.g., \citealt{Karas, geokerr, ynogk, ynogkm}) were written in {\tt Fortran},
many contemporary ray-tracing and radiation transfer codes (see, e.g., \citealt{Grmonty}; \citealt{Gyoto}; \citealt{Astroray}; \citealt{Ray}; \citealt{Ipole}; \citealt{Blacklight}) use {\tt C} or {\tt C++} for finer memory management and optimizations. \citet{Gray} developed the first radiative ray-tracing algorithm that makes use of general-purpose computing on graphics processing units (GPUs). The advent of GPU programming resulted in one to two orders of magnitude speed-up for relativistic ray-tracing codes (see, e.g., \citealt{Gray}). Since then, many GPU-based ray-tracing and radiation transfer codes have been developed (see, e.g., \citealt{Odyssey}; \citealt{Raptor}; \citealt{GRay2}; \citealt{Raptor2}), better enabling large-scale studies of black hole images.

Contemporary relativistic ray-tracing codes often achieve their remarkable speeds at the cost of decreased flexibility and user-friendliness. For example, the metric derivatives required for calculation in curved space-times are often hard-coded (cf.~e.g., \citealt{FANTASY}). However, ray-tracing in non-Kerr metrics has become an increasingly common method to help constrain the near-horizon black hole spacetime geometry with the Event Horizon Telescope (see e.g., \citealt{2020PhRvL.125n1104P,2021PhRvD.103j4047K,EHT_14} for gravitational tests and e.g. \citealt{Medeiros2020,Younsi2023} for simulations of non-Kerr metrics). Frequently, the user must manually calculate and implement a significant amount of new code to work with a new metric. This procedure is cumbersome, time consuming, and also error prone.

In this paper, we introduce {\tt Mahakala}\footnote{{\tt Mahakala} is named after an Indian deity mah$\bar{a}$k$\bar{a}$la believed to be the depiction of absolute black, and the one who has the power to dissolve time and space into himself, and exist as a void at the dissolution of the universe.}$^,$\footnote{\texttt{Mahakala} is open source and available on github at \href{https://github.com/liamedeiros/Mahakala}{https://github.com/liamedeiros/Mahakala}, see also \citet{sharma_2025_14985433}.}, a {\tt Python}-based, accelerated, ray-tracing and radiation transfer code for arbitrary space-times.\footnote{Here by arbitrary we mean that we do not assume stationarity or axisymmetry. However, we do assume that the geodesic equation still holds and that the metric is free of pathologies such as non-Lorentzian signatures and closed time-like loops (see \citealt{2013PhRvD..87l4017J} for a systematic study of pathologies in non-Kerr metrics).} 
Our aim with \texttt{Mahakala} is to balance speed with ease-of-use and flexibility: we have designed {\tt Mahakala} to be modular and portable so that it can make use of specialized hardware to run in parallel on graphics processing units (GPUs) and tensor processing units (TPUs) in addition to conventional central processing units (CPUs). However, since the code is written in {\tt Python}, it can also be easily run in a \texttt{jupyter notebook} on a laptop, lowering the barrier to entry into radiative ray-tracing. The modular nature of {\tt Mahakala} also allows the user to seamlessly use data from intermediate steps in the ray-tracing, e.g., to contrast the contribution of different regions of the flow or study the importance of different relativistic effects.

To parallelize mathematical operations, \texttt{Mahakala} uses {\tt JAX} (\citealt{jax2018github}), Google's new machine learning framework, which supports just-in-time (jit) compilation and vectorization. {\tt JAX} also provides an implementation of accelerated \emph{automatic differentiation}. 
Automatic differentiation stands in contrast to manual differentiation (which is cumbersome and error prone) and numerical differentiation (which is computationally expensive and can result in large numerical errors).
In automatic differentiation, a function is programmatically augmented to concurrently compute its derivative(s). This is achieved by decomposing the function into a graph rooted by base elementary operations whose derivatives are known (like additions and multiplications) and then recursively iterating through the graph, keeping track of the derivatives at each node, and applying the chain rule.
\texttt{Mahakala} uses automatic differentiation to compute Christoffel symbols directly from an input metric, so it can be easily and efficiently extended to work with non-Kerr geometries.

The paper is organized as follows. In Section~\ref{sec:algorithm}, we discuss the numerical schemes used by {\tt Mahakala} for ray-tracing calculations. Section~\ref{sec:RT_intro} reviews the equations of total intensity radiative transfer along with the synchrotron emissivity prescription used by {\tt Mahakala}. We illustrate the accuracy of the code with several tests in Section~\ref{sec:Tests}. In Section~\ref{sec:applications}, we analyze hundreds of snapshots from two {\tt AthenaK} GRMHD simulations with {\tt Mahakala} and demonstrate the algorithm's ability to determine where different image features originate in the flow. We summarize in Section~\ref{sec:conclusion}.

\section{{\tt Mahakala} Algorithm}\label{sec:algorithm}

\begin{figure}
    \includegraphics[width=1\columnwidth]{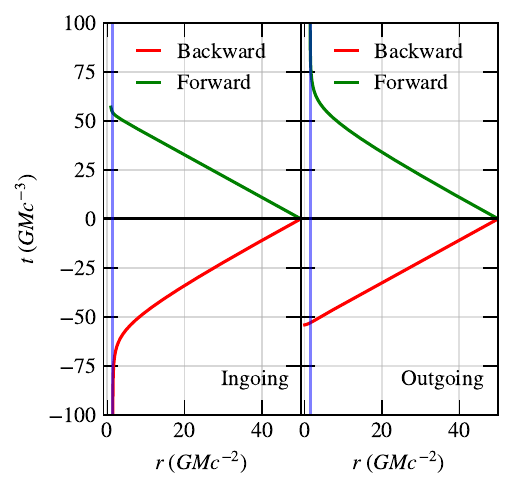}
    \caption{Comparison of photon trajectories through the Schwarzschild metric integrated both forwards \textit{(top)} and backwards \textit{(bottom)} in time, in both ingoing \textit{(left)} and outgoing \textit{(right)} Cartesian KS coordinates. In all panels the vertical blue line denotes the location of the event horizon. The ingoing Cartesian KS coordinates are horizon-penetrating for forward integration, whereas the outgoing Cartesian KS coordinates are horizon-penetrating for backward integration.
    }
    \label{fig:In_vs_Out}
\end{figure}

\begin{figure*}[hbt!]
    \centering
    \includegraphics[width=0.8\textwidth]{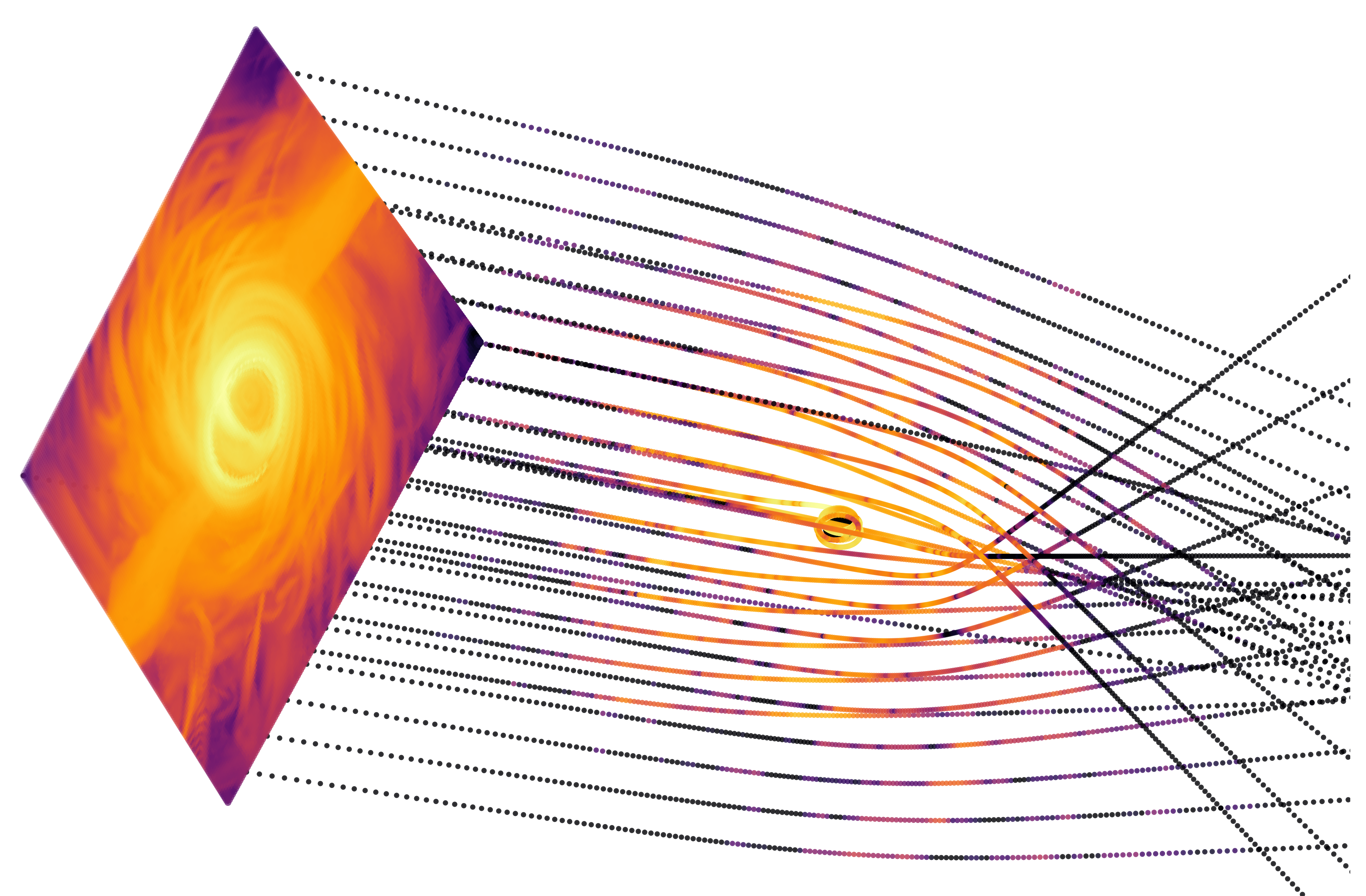}
    \caption{A visual representation of how {\tt Mahakala} simulates the trajectories of photons from the observer's image plane through the three dimensional space near a black hole. The image plane shows the resulting image with a logarithmic color scale. The color map along photon trajectories represents the total synchrotron emission at each point in the photon's path through the output of a GRMHD simulation.}
    \label{fig:3D_trajectories}
\end{figure*}

Many contemporary ray-tracing algorithms use either Boyer-Lindquist (BL) coordinates or spherical Kerr-Schild coordinates (see e.g., \citealt{Gold, 2023ApJ...950...35P} for recent reviews). However, when using a spherical coordinate system, the singularity at the pole ($\theta = 0, \, \pi$) can give rise to numerical errors. To avoid these issues, {\tt Mahakala} uses the Cartesian Kerr-Schild (KS) coordinate system. This choice of coordinates also allows us to seamlessly interface with the new \texttt{AthenaK} code (\citealt{2023ApJ...949..103W}, J.~Stone et al.~in preparation). In Cartesian KS coordinates, the Kerr metric is (see e.g., \citealt{Visser})
\begin{equation}
    g_{\alpha\beta} = \eta_{\alpha\beta} + fl_{\alpha}l_{\beta},
    \label{eq:metric}
\end{equation}
where $\eta_{\alpha\beta}$ =  $\mathrm{diag}(-1, 1, 1, 1)$ is the Minkowski metric, $f$ is given by 
\begin{equation}
    f = \frac{2Mr^3}{r^4 + a^2z^2},
    \label{eq:f}
\end{equation}
and 
\begin{equation}
    l_{\alpha} = \left(\pm 1,\frac{rx + ay}{r^2 + a^2},\frac{ry-ax}{r^2 + a^2},\frac{z}{r}\right).
    \label{eq:l_alpha}
\end{equation}
Here $M$ is the mass of the black hole, $a$ is the black hole spin parameter (i.e. the angular momentum of the black hole written in units of $M$) and $r$ is defined implicitly through
\begin{equation}
    x^2 + y^2 + z^2 = r^2 + a^2(1-z^2/r^2).
    \label{eq:r}
\end{equation} 

Throughout the paper we use the $\mathrm(-1, 1, 1, 1)$ metric signature and set $c=G=1$ unless otherwise stated, where $G$ is the gravitational constant, and $c$ is the speed of light. Greek indices vary from 0 to~3. The positive and negative signs in the $0$-th component of $l_{\alpha}$ correspond to the ingoing and outgoing Cartesian KS coordinates, respectively. In Figure~\ref{fig:In_vs_Out}, we compare the trajectories of photons through the Schwarzschild metric in both ingoing and outgoing Cartesian KS coordinates. As shown in the figure, photon trajectories integrated forward in time in ingoing Cartesian KS coordinates are horizon-penetrating, while trajectories integrated backwards in time approach the horizon asymptotically. For outgoing Cartesian KS coordinates, the opposite is true; photon trajectories are only horizon-penetrating if they are evolved backwards in time (see \citealt{2023PhRvD.108h4004B} for a detailed comparison of ingoing and outgoing KS coordinates). This result also holds for ingoing and outgoing spherical KS coordinates. 

{\tt Mahakala} can be used with both ingoing and outgoing Cartesian KS coordinates and integrates the photon trajectories backwards in time from the observer's image plane into the regions near the black hole. By default, {\tt Mahakala} uses the (non-horizon-penetrating) ingoing coordinates for consistency with the new \texttt{AthenaK} GRMHD code.

We follow the formalism of \citet{Johannsen_and_Psaltis} and initialize the observer's image plane at a distance $d$ from the black hole and at an inclination angle $i$ with respect to the black hole's spin axis (see Figure 1 in \citealt{Johannsen_and_Psaltis}). The center of the observer's image plane lies on the $x-y$ plane of the simulation coordinates. In Figure~\ref{fig:3D_trajectories} we show an example simulated image along with selected photon trajectories used to generate the image. We define $x'$ and $y'$ as the coordinates on the image plane and relate them to the Cartesian KS coordinates ($x,\,y,\,z$) as follows:
\begin{align}
    & x = -y'\cos i+d\sin i,\\
    & y = x',\\
    & z = y'\sin i + d\cos i.
    \label{eq:BH_transformations}
\end{align}
We initialize a photon at each pixel location such that the direction of its momentum is parallel to the vector connecting the center of the black hole to the center of the image plane. We normalize the photon's 4-momentum $k^{\mu}$ such that $k^0=1$ and $k_{\mu}k^{\mu} = 0$. We assume that the image plane is initialized at a large enough distance away from the black hole that the spacetime where the photons are initialized can be approximated as Minkowski. 

The geodesic equations in curved space-time are 
\begin{subequations}
    \begin{align}
        \frac{dx^{\mu}}{d\lambda} &= k^{\mu}, \\
        \frac{dk^{\mu}}{d\lambda} &= -\Gamma^{\mu}_{\alpha \beta} k^{\alpha} k^{\beta},
        \label{eq:geo_eq1}
    \end{align}
\end{subequations}
where $\lambda$ is an affine parameter and ${\Gamma^{\mu}}_{\alpha \beta}$ are the Christoffel symbols. We follow \citet{GRay2} and re-write equation~(\ref{eq:geo_eq1}) in terms of the metric derivative tensor,
\begin{equation}
    \ddot{x}^{\mu} = -\left(g^{\mu \beta}\dot{x}^{\alpha}-\frac{1}{2}g^{\mu \alpha}\dot{x}^{\beta}\right)g_{\beta \gamma,\alpha} \, \dot{x}^{\gamma}, 
    \label{eq:geo_eq3}
\end{equation}
where $\dot{x}^{\mu} \equiv dx^{\mu} \slash d\lambda$. In this form, the operation count of calculating the geodesic equation is significantly reduced,\footnote{It is computationally more expensive to solve the geodesic equation in Cartesian KS as compared to BL coordinates. However, rewriting the geodesic equation as equation~(\ref{eq:geo_eq3}) can reduce the computational cost (see, e.g., \citealt{GRay2}).} resulting in higher efficiency. We numerically integrate the geodesic equation backwards in time using a Runge-Kutta 4-th order (RK4) scheme. 

Most existing ray-tracing codes integrate the geodesic equation with respect to the affine parameter. However, using the affine parameter to integrate backwards in time through ingoing KS coordinates can lead to the accumulation of large error near the horizon due to the exponential growth of $k^0 \equiv dt \slash d\lambda$ (see \citealt{2023PhRvD.108h4004B}). Because of this, we include in \texttt{Mahakala} the ability to integrate with respect to either affine parameter or coordinate time, which avoids these errors. The geodesic equation in coordinate time can be written as 
\begin{equation}
    \frac{dv^i}{dt}=-\Gamma^{i}_{\alpha \beta}v^{\alpha}v^{\beta}+\Gamma^0_{\alpha \beta}v^{\alpha}v^{\beta}v^i,
    \label{eq:Geodesic_final}
\end{equation}
where $v^{i} \equiv dx^{i}/dt$ and $v^{0}= 1$. We again re-write the equation in terms of the metric derivative tensor
\begin{equation}
    \frac{dv^i}{dt}=-\left(g^{'i \beta}v^{\alpha} - \frac{1}{2}g^{'i \alpha} v^{\beta}\right)g_{\beta \gamma,\alpha}v^{\gamma},
    \label{eq:Final_eq}
\end{equation}
where $g^{'i\mu} \equiv g^{i \mu} - v^i g^{0\mu}$. {\tt Mahakala} integrates with respect to affine parameter by default.

\begin{figure}
    \includegraphics[width=\columnwidth]{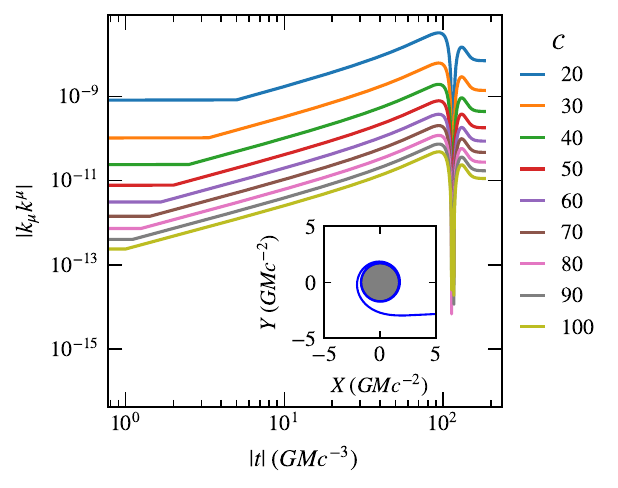}
    \caption{The value of $|k_{\mu}k^{\mu}|$, which is a measure of the error as a function of coordinate time for an example photon. Different colors correspond to different values of $\mathcal{C}$, and the inset shows the trajectory of the photon around a black hole with $a = 0.9\,M$. For efficiency, we choose $\mathcal{C}=50$ as the default value, but the user may specify a higher value if a smaller error is required.}
    \label{fig:Delta_vs_C}
\end{figure}

We use a semi-adaptive time-step for integration, where the step-size depends on the photon's location. The step-size $\mathcal{S}$ at iteration $i+1$ is given by 
\begin{equation}
    \mathcal{S}_{i+1} = \frac{r_i}{\mathcal{C}},
    \label{eq:step-size equation}
\end{equation}
where $r_i$ is the radial distance of the photon from the center of the black hole\footnote{When integrating the geodesic equation in affine parameter ($\lambda$), $r_i$ is defined as the distance of the photon from the horizon.} for the $i$-th iteration and $\mathcal{C}$ is a free parameter that scales the step size. {\tt Mahakala} does not explicitly use constants of motion to integrate the geodesic equation. Because of this, $k_{\mu} k^{\mu} \equiv k^2$ is unconstrained and is a good measure of the numerical error. Figure~\ref{fig:Delta_vs_C} shows $|k_{\mu}k^{\mu}|$ as a function of time for different values of $\mathcal{C}$. As the value of $\mathcal{C}$ is increased from 20 to 50, the error decreases significantly from  $\sim10^{-8}$ to $\sim10^{-10}$. Since increasing $\mathcal{C}$ further has a relatively small effect on the error and to ensure computational efficiency, we set $\mathcal{C}=50$ by default.\footnote{The analysis in Figure~\ref{fig:Delta_vs_C} is for integration with respect to coordinate-time. When performing integration with respect to affine parameter, we use $\mathcal{C}=100$ as the default value for computational efficiency.}

Equations~(\ref{eq:Geodesic_final}) and (\ref{eq:Final_eq}) are only valid where $dt \slash d{\lambda}$ is finite and well-defined. For integration backwards in time in ingoing Cartesian KS coordinates, $dt \slash d\lambda$ grows exponentially near the horizon and becomes infinite at the horizon. To avoid this, {\tt Mahakala} stops the integration for photons that approach within a distance of $\epsilon=10^{-4}\,M$ from the horizon. We also stop integrating the photons that reach distances larger than $d$, the distance between the black hole and the image plane. Results shown here use the default value of $d=1,000M$.

\section{Synthetic Images of GRMHD Simulations}\label{sec:RT_intro}

Our primary goal with {\tt Mahakala} is to simulate the mm-wavelength emission of low-luminosity accretion flows such as the ones onto the galactic center black hole, Sagittarius~A$^*$ (Sgr~A$^*$), and the supermassive black hole in M87. {\tt Mahakala} calculates synchrotron emissivity since it is the primary emission mechanism for these sources at these wavelengths (see, e.g., \citealt{Genzel_2010}). To demonstrate {\tt Mahakala}'s capabilities, we simulate 1.3~mm images of snapshots from GRMHD simulations performed with the new \texttt{AthenaK} code (\citealt{Athena++}, J.~Stone et al. in preparation). We use a nested mesh uniform Cartesian grid and linearly interpolate between grid points to calculate the values of the primitive variables (the variables natively output by the GRMHD simulations) along geodesic trajectories. We use GRMHD simulations that do not account for radiative effects like pressure or cooling, since the effects of radiation on the dynamics of the flow are negligible for M87 and Sgr~A$^*$. Throughout the manuscript, we also assume that the GRMHD flow does not change in the time it takes for the photon to move across the simulation domain, i.e. we adopt the fast-light approximation. 

The primitive variables of the {\tt AthenaK} simulations are the fluid-frame density $\rho$, fluid-frame gas pressure $p_{\text{gas}}$, spatial components of the fluid velocity in the normal-frame\footnote{The frame that is orthogonal to surfaces of constant coordinate time.} $u'^{i}$, and the spatial components of the magnetic field in Cartesian coordinate frame $B^{i}$. The contravariant components of the magnetic field measured by the fluid and expressed in the coordinate frame $b^{\mu}$ are given by (see, e.g., \citealt{White_et_al_2016, Athena++})
\begin{subequations}
    \begin{align*}
    b^{0} &= u_iB^{i}, \\
    b^i &= \frac{1}{u^0}(B^{i} + b^{0}u^{i}),
    \label{eq:Primitive_B_transform}
    \end{align*}
\end{subequations}
where $u^0 = \gamma / \alpha$, $u^i = u'^{i} - \beta^{i} \gamma / \alpha$, $\alpha = ( - g^{00})^{-1/2}$ is the lapse, $\beta^{i} = \alpha^2g^{0i}$ is the shift, and 
\begin{equation}
    \gamma = (1 + g_{ij} u'^{i} u'^{j})^{1/2}
    \label{eq:Primitive_U_tranform}
\end{equation}
is the Lorentz factor in the normal frame. We denote the contravariant components of the fluid velocity in the coordinate frame as $u^{\mu}$ throughout. We interpolate the primitive variables rather than $u^{\mu}$ and $b^{\mu}$ to ensure that $u_{\mu}u^{\mu} = -1$ and $u_{\mu}b^{\mu} = 0$.

\subsection{Radiative Transfer}\label{subsec:RT}

Accounting for emission and absorption (and neglecting scattering), the covariant form of the general relativistic radiative transfer equation for total intensity is (see, e.g., \citealt{Ziri_Younsi})
\begin{equation}
    \frac{d\mathcal{I}}{d\lambda}=-k^{\mu} u_{\mu} \left(-\alpha_{\nu,0}\mathcal{I} + \frac{j_{\nu,0}}{\nu^3}\right).
    \label{eq:RT}
\end{equation}
Here $\mathcal{I}=I_{\nu}/\nu^3$ is Lorenz-invariant and $I_{\nu}$ is the specific intensity of the ray at frequency $\nu$. In the above equation, $\alpha_{\nu}$ and $j_{\nu}$ correspond to the absorption and synchrotron emissivity at frequency $\nu$, respectively. Quantities with subscript $0$ are evaluated in the local frame of the plasma.  The frequency of radiation measured by an observer with four-velocity $u^\mu$ is
\begin{equation}
    \nu = -k^{\mu}u_{\mu},
    \label{eq:nu_def}
\end{equation}
where $k^{\mu}$ is the contravariant 4-momentum of the photon.

The modular nature of {\tt Mahakala} also allows us to calculate the radiative transfer equation~(\ref{eq:RT}) in a separate step, which can result in a significant speed up when generating multiple images from the same set of geodesics---such as when varying parameters that only affect the radiative transfer (e.g., $n_e$, $\mathcal{M}$, and $R_{\mathrm{high}}$ defined below) or for different simulations with fixed spacetime and observer inclination. 

As discussed in section~\ref{sec:algorithm}, {\tt Mahakala} can integrate the geodesic equation with respect to either the affine parameter or coordinate time. To solve the radiation transfer equation when integrating in coordinate time, we use the chain rule to write equation~(\ref{eq:RT}) as 
\begin{equation}
    \frac{d\mathcal{I}}{dt}=-k^{\mu} u_{\mu} \left(-\alpha_{\nu,0}\mathcal{I} + \frac{j_{\nu,0}}{\nu^3}\right)\kappa,
    \label{eq:RT_new}
\end{equation}
where $\kappa = d\lambda/dt$. We calculate $\kappa$ by solving the following pair of coupled differential equations
\begin{align}
    \frac{d\lambda}{dt} &= \kappa, 
    \label{eq:kappa_1} \\
    \frac{d\kappa}{dt} &= \kappa {\Gamma^0}_{\alpha \beta}v^{\alpha}v^{\beta},
    \label{eq:kappa_2}
\end{align}
where $v^{\mu} = dx^{\mu}/dt$. We decrease the computational expense of solving equation~(\ref{eq:kappa_2}) by writing it in terms of the metric derivative tensor as done for equation~(\ref{eq:geo_eq3}).

\subsection{Emissivity}\label{subsec:Simulated_emission}
We restore $c$ throughout section~\ref{subsec:Simulated_emission}. We adopt the following approximate expression for thermal synchrotron emissivity \citep{Po_Kin_leung}
\begin{equation}
    j_{\nu} = n_e \frac{\sqrt{2} \pi e^2 \nu_s}{3 K_2(1/\Theta_e) c} (X^{1/2} + 2^{11/12}X^{1/6})^2 \exp{(-X^{1/3})}.
    \label{eq:Synchrotron_emission}
\end{equation}
Here $e$ is the electron charge, $n_e$ is the electron density, and $K_2$ is the modified Bessel function of the second kind for integer order 2, and
\begin{equation}
    X = \frac{\nu}{\nu_s},
    \label{eq:X}
\end{equation}
where $\nu_s = (2/9)\nu_c \Theta_e^2 \sin \theta_B$, and the cyclotron frequency $\nu_c$ is given by
\begin{equation}
    \nu_c = \frac{e B}{2\pi m_e c}.
    \label{eq:nu_s}
\end{equation}
The pitch angle $\theta_B$ is the angle between the emitted or absorbed photon and the magnetic field vector $B$ as evaluated in the fluid frame
\begin{equation}
    \cos \theta_B = \frac{k_{\mu}b^{\mu}}{-k_{\mu}u^{\mu}\sqrt{b_{\mu}b^{\mu}}}.
    \label{eq:Pitch_angle}
\end{equation}
The dimensionless electron temperature is
\begin{equation}
    \Theta_e \equiv \frac{kT_e}{m_e c^2},
    \label{eq:electron_temp}
\end{equation}
where $k$ is the Boltzmann constant, $T_e$ is the electron temperature, and $m_e$ is the mass of the electron. We calculate the absorption coefficient using Kirchoff's Law (see, e.g., chapter 1 of \citealt{Rybicki_Lightman}).

Low-luminosity accreting black holes, such as the ones in M87 and Sgr~A$^*$, are expected to have advection dominated accretion flows (ADAFs; see, e.g., \citealt{ADAF_Narayan} for a review). Due to their low accretion rates, ADAFs have such low densities that they are effectively Coulomb collisionless. As a result, their ions and electrons may not reach thermal equilibrium, thus producing a two-temperature plasma (see, e.g., \citealt{Eliot,Eliot_and_Gruzinov}). Despite this, and because the electrons likely contribute negligibly to the overall flow energetics, many contemporary GRMHD simulations only evolve a single plasma temperature or internal energy (although see e.g., \citealt{Sean_2015, Andrew_2_temp}).  
To recover the electron temperature from the GRMHD variables, we set the electron-to-ion temperature ratio ($T_i/T_e$) based on the local ratio of gas to magnetic pressure in the plasma, $\beta = p_{\text{gas}}/p_{\text{mag}}$, as follows (see, e.g., \citealt{Moscibrodzka,M87_paper_5}),
\begin{equation}
    \frac{T_i}{T_e} = R_{\mathrm{high}}\frac{\beta ^2}{1 + \beta ^2} + \frac{1}{1+\beta^2},
    \label{eq:Ti/Te}
\end{equation}
where $R_\mathrm{high}$ is a free parameter. Following \citet{Patoka}, we do not set the ion temperature equal to the fluid (gas) temperature to avoid overcounting the energy in the system. Instead, we calculate $T_e$ using the total fluid internal energy, 
\begin{equation}
    T_e = \frac{m_p u (\hat{\gamma_e} -1) (\hat{\gamma_i} - 1)}{k \rho ((\hat{\gamma_i}-1) y + (\hat{\gamma_e}-1) Rz)},
    \label{eq:Te}
\end{equation}
where $m_p$ is the proton mass and $u$ is the internal energy. 
Here $1/y$ and $1/z$ correspond to the total number of electrons and nucleons per atom, with $y=z=1$ for pure hydrogen. We set the adiabatic index for the ions to $\hat{\gamma_i} = 5/3$ since they are typically non-relativistic and $\hat{\gamma_e} = 4/3$ for the typically relativistic electrons.

\section{Tests from the Literature}\label{sec:Tests}
In this section we ensure the accuracy of {\tt Mahakala} by performing several tests from the literature. Most tests included in this section are reproduced from \citet{Gold}, where a large number of radiative ray-tracing codes were compared against each other.


\subsection {Null Geodesic Deflection} \label{sec:Deflection_test}
We begin with a test of how the azimuthal deflection angle of null geodesic trajectories ($\Delta \phi$) depends on the impact parameter $b$ \citep{Gold}. Analytic solutions for the deflection angle of null geodesics confined to the $x-y$ plane in the Kerr metric can be obtained by a quadrature of standard, elliptic functions given in \citet{Iyer_and_Hansen}. For this test, we uniformly vary the impact parameter $b$ from $-20\,M$ to $20\,M$ in intervals of $0.4\,M$. We ignore photons with impact parameter satisfying $b_-\leq b \leq b_+$ as these photons fall inside the black hole. Here, $b_-$ and $b_+$ are given by
\begin{equation}
    b_{\pm} = -a \pm 6M\cos{\bigg(\frac{1}{3}\cos^{-1}\bigg(\mp\frac{a}{M}\bigg)\bigg)}.
    \label{eq:crit_impact_param}
\end{equation}

\begin{figure}
    \includegraphics[width=1\columnwidth]{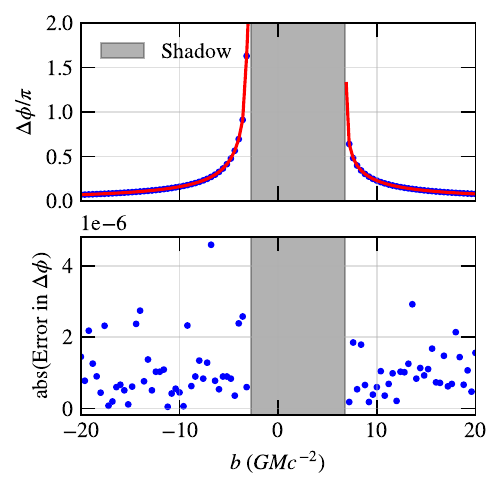}
    \caption{(\textit{top}) Deflection angle of null geodesics in the equatorial plane around a black hole with $a=0.9\,M$ integrated with {\tt Mahakala}, as a function of the impact parameter~$b$. (\textit{bottom}) The absolute error, as compared to analytic solutions, in the deflection angle calculated by {\tt Mahakala}.}
    \label{fig:Deflection_angle}
\end{figure}

We show that the numerical results from {\tt Mahakala} show good agreement with the analytic formula in Figure~\ref{fig:Deflection_angle}, where we have set the black hole spin parameter to $a=0.9\,M$ and the distance to the black hole from the observer's image plane to $d=1000\,M$. The absolute error between the numerical and analytical results (lower panel of Figure~\ref{fig:Deflection_angle}) is of the order of $~10^{-6}$. Our range of errors is consistent with the other codes included in \citet{Gold}. 
\begin{figure}
    \includegraphics[width=1\columnwidth]{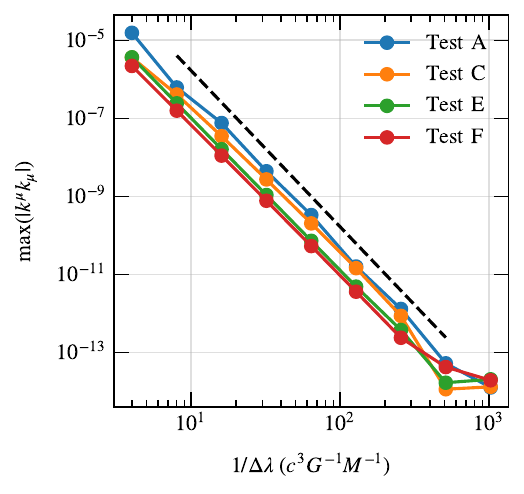}
    \caption{Maximum error, $\mathrm{max}(|k \cdot k|)$, as a function of the inverse of step size, $1 \slash \Delta \lambda$, for the four spherical orbit tests shown in Figure~\ref{fig:CT_all}. We vary the step size from $1/4\,M$ to $1/1024\,M$ in factors of two. {\tt Mahakala}'s geodesic integration scheme converges at 4--th order, i.e., $y\sim x^{-4}$ (black dashed line) as expected for an RK4 method.}
    \label{fig:Convergence_test_fig_5}
\end{figure}
\subsection {Unstable Spherical Photon Orbits } \label{sec:Convergnece_1}
\begin{figure*}[hbt!]
    \centering
    \includegraphics[width=0.8\textwidth, height = 0.8\textwidth]{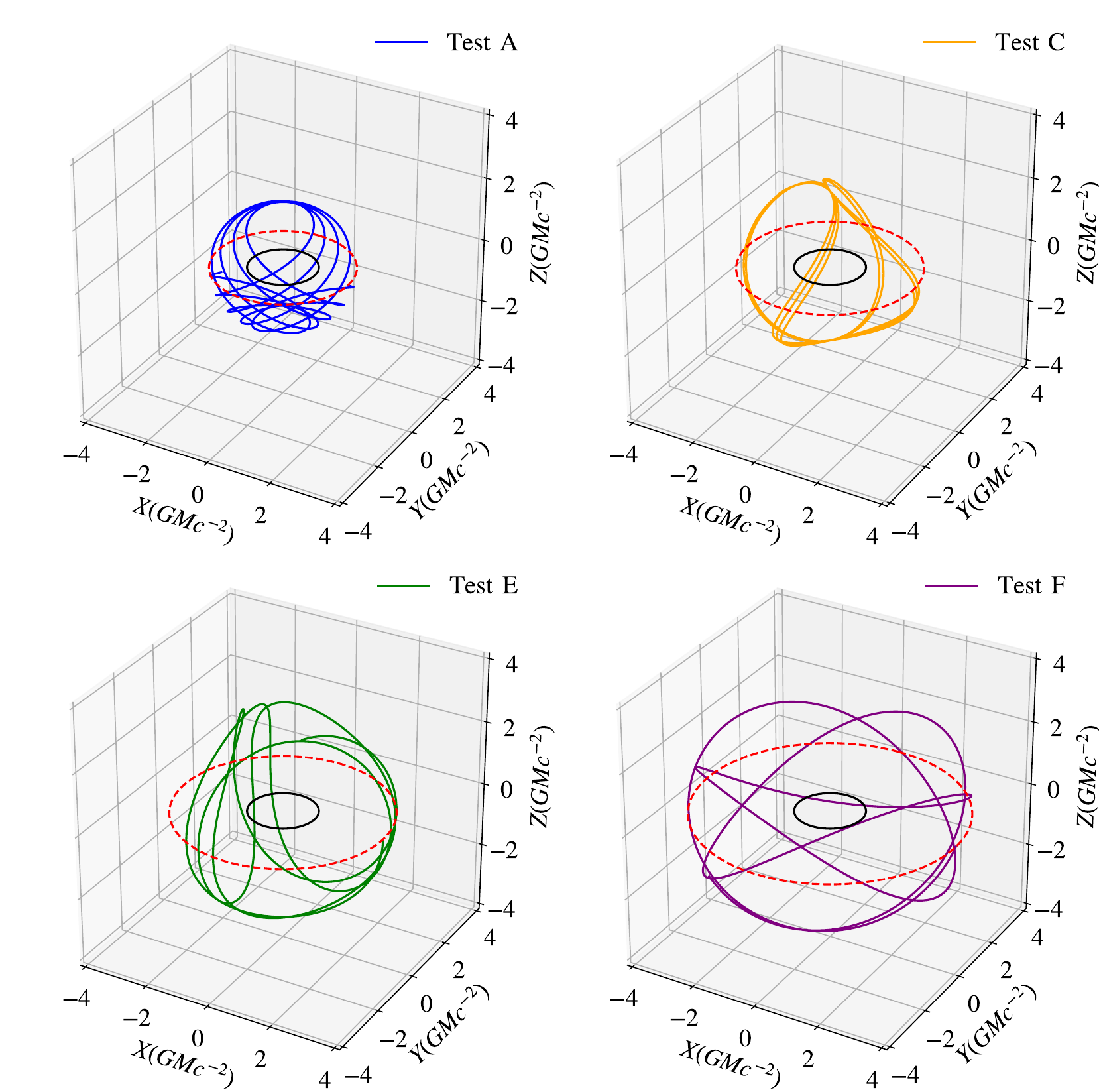}
    \caption{Unstable spherical photon orbits around a black hole with spin $a=M$ and time step $\Delta \lambda = 1 \slash 1024\,M$. For all panels, solid colored lines are the photon orbits, black solid circles are the (physical) singularities, and red dashed circles mark the radii of the spherical orbits on the equatorial planes. All orbits start by moving upward from the positive x-sides of the red dashed circles, and are integrated forwards in time for a fixed number of iterations.}
    \label{fig:CT_all}
\end{figure*}

Here we test the accuracy of {\tt Mahakala}'s RK4 geodesic integration with convergence tests for unstable spherical photon orbits (see, e.g., \citealt{GRay2}). Integration of spherical photon orbits allows us to test the long term behavior of our algorithm since the errors will accumulate along these trajectories. We set $a=M$ and use a constant time-step for this test.

For a black hole with mass $M$ and spin $a$, spherical photon orbits will lie between the prograde radius ($r_p$), and the retrograde radius ($r_r$),
\begin{align}
    r_p &= 2M \bigg\{1 + \cos\bigg[\frac{2}{3}\cos^{-1}\bigg(-\frac{|a|}{M}\bigg)\bigg]\bigg\},
    \label{eq:r_p} \\
    r_r &= 2M \bigg\{1 + \cos\bigg[\frac{2}{3}\cos^{-1}\bigg(\frac{|a|}{M}\bigg)\bigg]\bigg\},
    \label{eq:r_r}
\end{align}
where $r_p$ and $r_r$ also satisfy the inequality $M \leq r_p \leq 3M \leq r_r \leq 4M$. {\tt Mahakala} then calculates the normalized angular momentum $\Phi$ and the Carter constant $Q$ used to identify the initial conditions for the photons (see, e.g., \citealt{Teo_2003}),
\begin{align}
    \Phi &=- \frac{r^3 - 3Mr^2 + a^2r + a^2M}{a(r-M)},
    \label{eq:PHI} \\
    Q &= -\frac{r^3(r^3 - 6Mr^2 + 9M^2r- 4a^2M)}{a^2(r-M)^2}.
    \label{eq:Q}
\end{align}

In Figure~\ref{fig:Convergence_test_fig_5} we demonstrate that {\tt Mahakala} converges as expected for RK4 integration with convergence plots for the four spherical photon orbits\footnote{These are orbits $\mathrm{A}$, $\mathrm{C}$, $\mathrm{E}$ and $\mathrm{F}$ from Table~1 in \citet{GRay2}.} considered in \citet{GRay2}. 
Figure~\ref{fig:CT_all} shows the results of simulating these trajectories with {\tt Mahakala}. The photon trajectories remain stable for several orbits and are consistent with the results of \citet{GRay2}.

\begin{figure*}[hbt!]
    \centering
    \includegraphics[width=\textwidth]{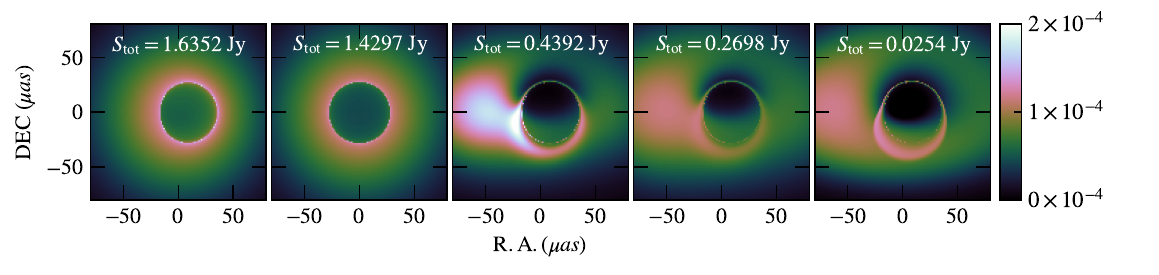}
    \caption{Images produced by {\tt Mahakala} for the five parameterized models in \citet{Gold}. The image morphology and total flux are consistent with the results of the several radiative ray-tracing algorithms in that paper. The colormap corresponds to $S/S_{\mathrm{tot, exact}}$ as in \citet{Gold}.}
    \label{fig:Gold_image_test}
\end{figure*}

\begin{figure*}[hbt!]
    \centering
    \includegraphics[width=\textwidth]{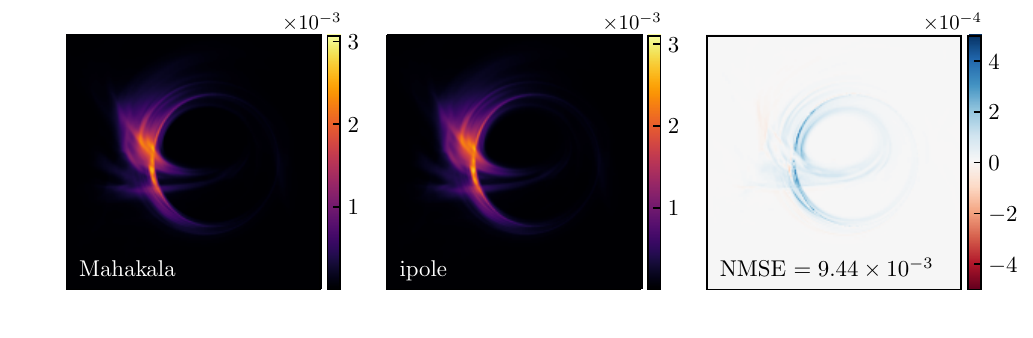}
    \caption{Comparison between the images generated by \texttt{Mahakala} (\textit{left}) and the widely used radiative ray tracing code \texttt{ipole} (\textit{middle}). The rightmost panel shows the difference between the two images and reports the NMSE (see equation~\ref{eq:NMSE} and \citealt{2023ApJ...950...35P}). The NMSE between the images is $9.44\times 10^{-3}$, which is well below the reported maximum, $0.02$, between the ray-tracing codes compared in \citet{2023ApJ...950...35P}.  
    }
    \label{fig:comparison_ipole}
\end{figure*}

\begin{figure*}
    \centering
    \includegraphics[width=\textwidth]{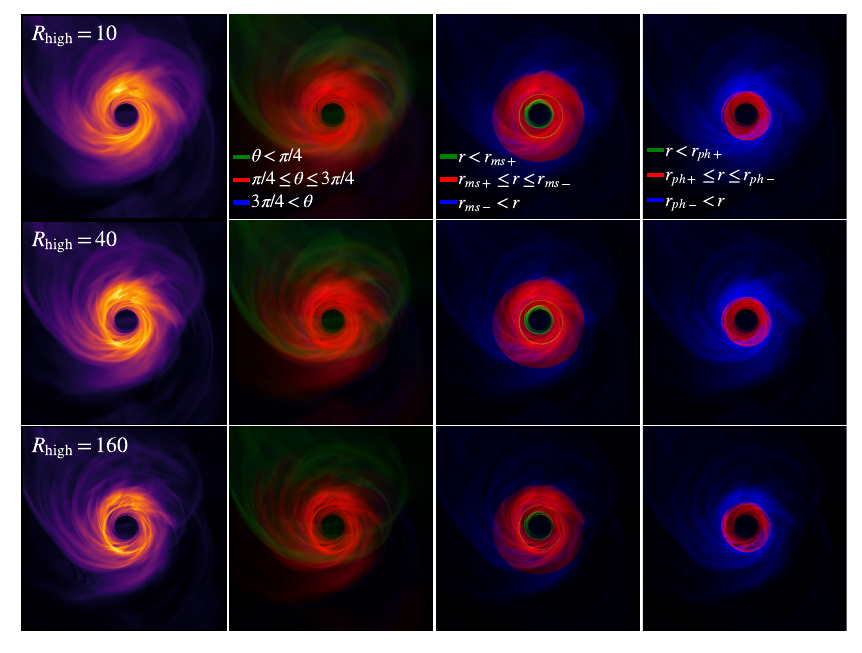}
    \caption{({\it first column}) Example 1.3~mm image simulated with {\tt Mahakala} for a MAD snapshot with spin $a = 0.9375\,M$,inclination $i=17^{\circ}$, electron temperature prescription set by $R_{\mathrm{high}}=10,\, 40,\, \mathrm{and} \, 160$ (top, middle, and bottom rows respectively), and a mass density scale of $\mathcal{M} = 5\times10^{25}$. ({\it second column}) The same snapshot but with color denoting the region where the emission was produced. Green denotes emission from the forward jet ($\theta<\pi/4$); red denotes emission from the disk region ($\pi/4\le \theta\le 3\pi/4$); and blue denotes the counter jet ($3\pi/4<\theta$). 
    The final color of each pixel is determined by the relative contribution of emission from each region.
    The third and fourth columns show 3-color images where each color corresponds to a spherical shell in the 3-D flow. We define $r_{ms\pm}$ and $r_{ph\pm}$ as the innermost stable circular orbit (ISCO) and photon orbit, respectively (see, e.g., \citealt{Bardeen_Press_Teukolsky}), where plus corresponds to the prograde orbits and minus to retrograde. For this black hole spin, the prograde and retrograde ISCO are at 2.04 $M$ and 8.82 $M$. The prograde and retrograde photon orbits are at 1.43 $M$ and 3.94 $M$. We use a gamma semi-log scale for all panels with $\Gamma=0.4$ (i.e., the colors show intensity to the $\Gamma$ power). The low-intensity ridges that can be seen in some images are artifacts due to the GRMHD simulation zones.}
    \label{fig:3_color_panel}
\end{figure*}

\begin{figure*}
    \centering
    \includegraphics[width=1\textwidth]{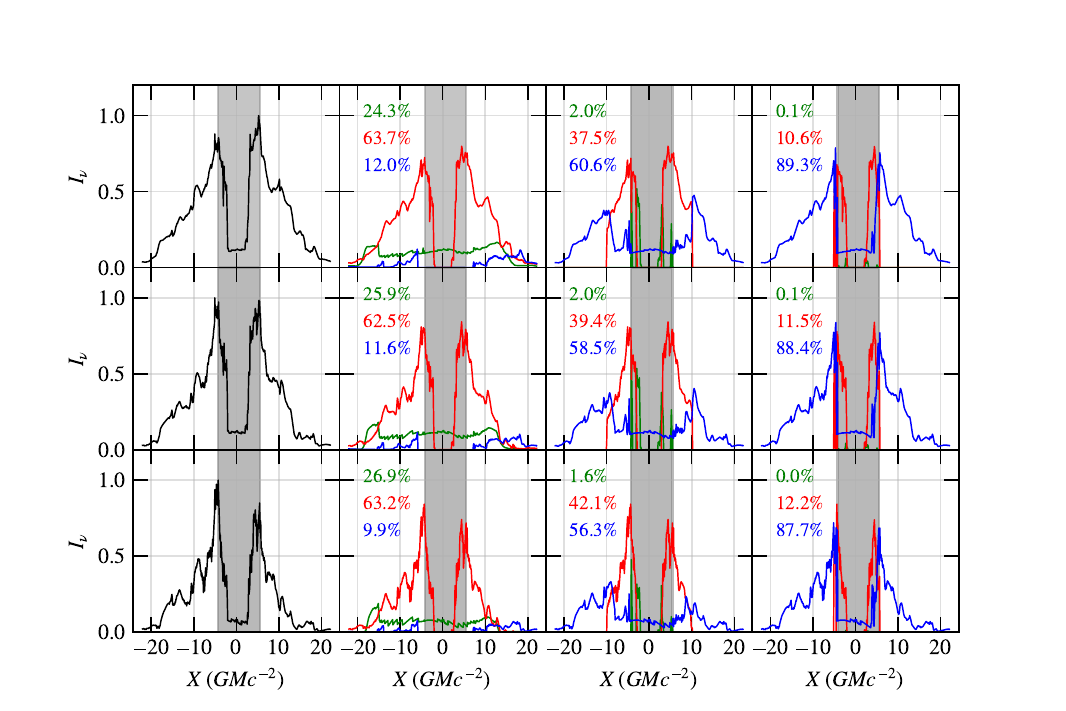}
    \caption{Horizontal cross sections of the total intensity and three color images in Figure~\ref{fig:3_color_panel}. The colors of the curves in columns $2-4$ are the same as those in Figure~\ref{fig:3_color_panel} and the percentages quoted at the top left of each panel correspond to the percentage of the total emission of the image that originates from each flow region. The shaded gray region shows the black hole shadow, i.e., the region within which all photon trajectories fall into the black hole.} 
    \label{fig:cross_sect_MAD}
\end{figure*}

\subsection{Images from Analytic Models} \label{sec:Image_test}

We test the radiative transfer components of {\tt Mahakala} by reproducing the analytic image test described in section 3.2 of \citet{Gold}. The test comprises an analytic accretion model simulated without the effects of scattering or polarization. We repeat this test with the same initial parameters, camera position, and parameter values as in \citet[][see their Table~1 for parameter values]{Gold}. We show the resulting images in Figure~\ref{fig:Gold_image_test}. The image morphology and total flux values are consistent with those of \citet[][see their Table~2 for total flux values]{Gold}.

\subsection{Normalized Mean Squared Error} \label{sec:NMSE}
We compare \texttt{Mahakala} to the open source radiative transfer and ray-tracing code \texttt{ipole} \citep{Ipole,Patoka}. We compare images produced by the two codes following the procedure described in \citet{2023ApJ...950...35P}. We calculate the normalized mean squared error (NMSE) between the two images, 
\begin{equation}
\mathrm{NMSE}(A,B) = \frac{\sum_j |A_j-B_j|^2}{\sum_j |A_j|^2},
\label{eq:NMSE}
\end{equation}
where $A_j$ is the intensity of the \texttt{ipole} image at pixel $j$ and $B_j$ is the intensity of the \texttt{Mahakala} image at the same pixel $j$. We compare the two simulated images and their NMSE in Figure~\ref{fig:comparison_ipole}. The total NMSE between the images is $\mathrm{NMSE}=9.44\times10^{-3}$, which is well within the range quoted for similar codes in \citet[see, e.g., the top right of their Figure~7]{2023ApJ...950...35P}.

\section{Applications}\label{sec:applications}

\begin{figure*}
    \centering
    \includegraphics[width=1\textwidth, height = 0.66\textwidth]{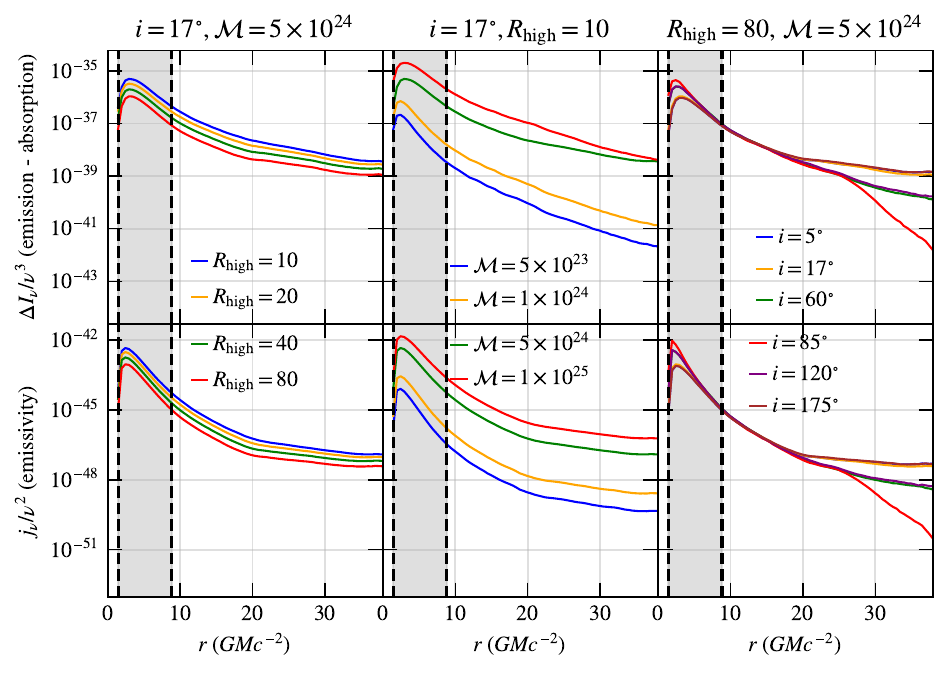}
    \caption{Average specific intensity ({\it top row}) and invariant synchrotron emission ({\it bottom row}) as a function of radius for the MAD simulation, averaged over $5000\,M$ 50 snapshots). In the left panels we vary $R_{\mathrm{high}}$ and set $\mathcal{M} = 5\times 10^{24}$ and $i=17^{\circ}$. In the middle panels we vary $\mathcal{M}$ and set $R_{\mathrm{high}}=10$ and $ i=17^{\circ}$. In the right panels we vary $i$ and set $R_{\mathrm{high}}=80$ and $\mathcal{M} = 5\times 10^{24}$. The vertical black dashed lines  correspond to the retrograde ISCO ($8.82\,M$) and the prograde photon orbit ($1.43\,M$). In all cases the emission peaks close to the prograde photon orbit, and the majority of emission originates from within the retrograde ISCO.}
    \label{fig:Changing_R_high_MAD}
\end{figure*}
\begin{figure*}
    \centering
    \includegraphics[width=1\textwidth, height = 0.66\textwidth]{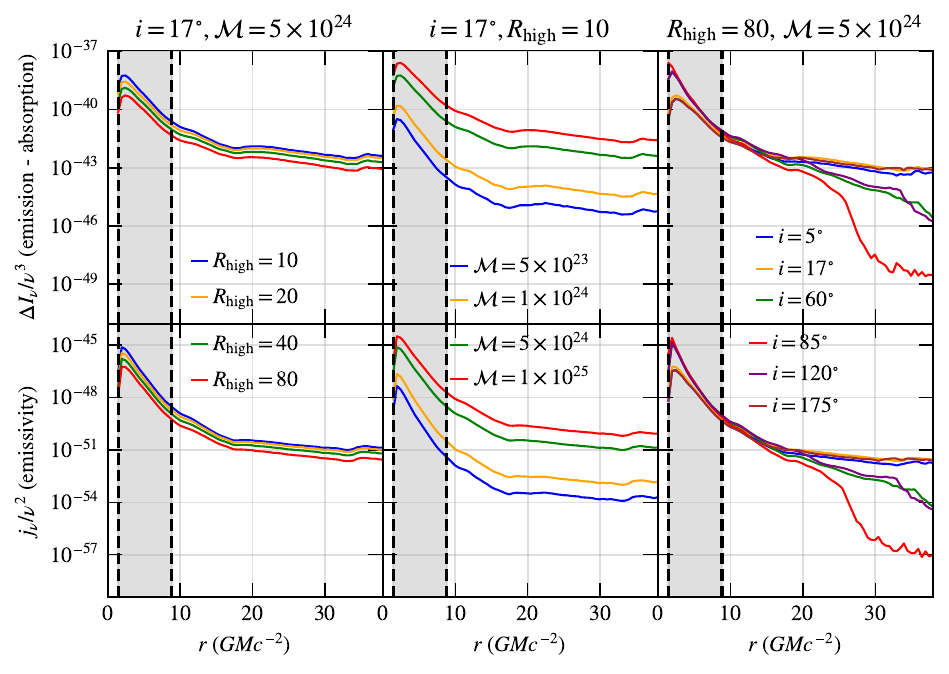}
    \caption{Same as Figure \ref{fig:Changing_R_high_MAD} but for a SANE simulation. All other parameters remain the same.}
    \label{fig:Changing_R_high_SANE}
\end{figure*}

{\tt Mahakala} is written with flexibility and ease of use as a high priority. Its modular design allows for efficient exploration of the relationship between image features and different parts of the accretion flow. It also enables the user to probe the importance of the physical mechanisms that determine the emission and propagation of light through the fluid domain. The ability to heuristically and quantitatively decompose an image in this way has many benefits, including:

\vspace{0.3em}

\noindent
\textit{(i)}
{\tt Mahakala} can be used to compute the amount of emission coming from regions of inflow/outflow or from the disk/jet (or counterjet). Different parts of the flow have different plasma properties and are subject to different heating mechanisms, so the ability to identify the origin of emission is crucial when attempting to connect observations to underlying physics. Emission from different locations will also undergo different general relativistic effects as it propagates to the observer. Understanding and constraining the geometric origin of emission therefore also helps improve our understanding of the robustness of tests of general relativity from horizon-scale images.

\vspace{0.3em}

\noindent
\textit{(ii)}
{\tt Mahakala}'s design also enables a quantitative comparison of the importance of different physical effects on the image features accessible to the EHT. One of the primary science outputs of EHT observations has been the image brightness asymmetry and structure \citep[][see also \citealt{2022ApJ...924...46M}]{M87_paper_5,SgrA_paper_5}. Doppler beaming, gravitational redshift, absorption through the disk, and the underlying radial emission profile all influence the ring brightness morphology. It is therefore important to quantitatively categorize the importance of these effects in different accretion models in order to interpret the observational data.

\vspace{1em}

We now use {\tt Mahakala} to simulate 1.3~mm images of GRMHD snapshots produced by the new {\tt AthenaK} code (\citealt{Athena++,AthenaK}). We consider two nested-mesh Cartesian grid GRMHD simulations with black hole spin $a=0.9375\,M$. 

The simulations were initialized to produce either the magnetically arrested disk (high magnetic flux, MAD; \citealt{2012MNRAS.426.3241N}) or standard and normal evolution (low magnetic flux, SANE; \citealt{Igumenshchev}) steady state configuration. Each simulation was run until $t = 10,000\,M$ in a grid extending to $\pm 1024\,M$ in each of the $x$, $y$, and $z$ directions. Each level of refinement was initialized with a resolution of $128$ grid zones across, and the innermost refinement level had a resolution of $16$ zones per $M$ spanning between $\pm 4\,M$ in $x$, $y$, and $z$.
We simulate images of 100 GRMHD snapshots (50 MAD and 50 SANE; over the last $5,000\,M$ of each simulation) with a time resolution of $100\,M$. Throughout this section we use ray-traced images with a field of view of $64\,M$ and an image resolution such that the pixel side length is $1/8 \,M$ (see \citealt{Psaltis_et_al_Resolution} for an exploration of the effects of resolution in simulated images). 

The equations of GRMHD are invariant under rescalings of length (or equivalently time$^{-1}$) and mass; however, the radiative transfer equations depend on both length (which is set by the black hole mass) and density (or equivalently the accretion rate in physical units). We select a black hole mass of $M = 6.5 \times 10^9 M_\odot$ for consistency with the supermassive black hole in M87 (see \citealt{Gebhardt_et_al,EHT_paper_6}). We parameterize the density scale normalization as $\mathcal{M}/\mathcal{L}^3$, where $\mathcal{L}=GMc^{-2}$ is the length scale (here we have restored the gravitational constant and the speed of light for clarity; see also \citealt{Patoka}). The units of $\mathcal{M}$ are such that multiplying the density from the fluid simulation by $\mathcal{M}/\mathcal{L}^3$ yields a value in g/cm$^3$. 

For simplicity, we will not include units when quoting values of $\mathcal{M}$, as they depend on the (arbitrary) code units used in the fluid simulation. For reference, a code mass unit of $\mathcal{M} = 5 \times 10^{25}$ corresponds to a mass accretion rate of $\dot{m} = \dot{M}/\dot{M}_{\rm Edd} = 2.8 \times 10^{-6}$ for the MAD model and $\dot{m} = 2.7 \times 10^{-7}$ for the SANE model.
The mass density scale is used to determine physical units for the total electron density, internal energy, and magnetic field strength. In contrast to the simulation library considered in \citet{M87_paper_5,SgrA_paper_5}, we do not use the total flux at 1.3~mm to set the mass scale $\mathcal{M}$, but rather vary it independently from the other free parameters. This allows us to explore the effect of the accretion rate independent of the other variables.

\vspace{1em}

As an example, we calculate the relative contribution from different flow regions in Figure~\ref{fig:3_color_panel}. The first column on the left shows the image of a GRMHD snapshot from a MAD simulation observed at 17 degree inclination, which is consistent with the inclination angle inferred for the large scale jet observed at radio frequencies \citep{2018ApJ...855..128W}. Each row in the figure corresponds to different values of $R_{\mathrm{high}}$. The second column shows three-color images with each color tracing emission from a different region of space (the forward jet in green, counter jet in blue, and disk regions in red). In computing the values for each color, we ignore absorption outside of the trial emission region.

Since the color perception of the human eye is not uniform, we also show the horizontal cross sections of these three color images in Figure~\ref{fig:cross_sect_MAD} and quote the percentage of flux that originates in each region. The disk contributes the most to the image, followed by the forward and then counter jets respectively. Specifically, the ring-like feature in the image is dominated by emission that originates within the disk. The disk emission peaks approximately at the edge of the black hole shadow, the boundary of which is defined as the critical impact parameter between the photons that escape and those that fall into the black hole. The forward jet contributes to the region above the ring and to the center of the ring, since at this inclination this emission originates between the black hole and the observer.

The third columns of both Figures~\ref{fig:3_color_panel} and \ref{fig:cross_sect_MAD} also show three color images, but with the different colors corresponding to spherical shells (with boundaries defined by the prograde and retrograde ISCO radii; see figure caption). In this example, matter between the prograde and retrograde ISCO radii contributes the most to the broad ring-like feature for all values of $R_{\mathrm{high}}$, while the region outside of the retrograde ISCO radius contributes the most to the overall flux of the image. We also observe a thinner high intensity ring at the edge of the black hole shadow, which originates primarily from within the prograde ISCO. The region within the prograde ISCO also contributes emission to a feature that peaks at the boundary of the ``inner shadow'' (see \citealt{2021ApJ...918....6C}). 

The fourth column is similar to the third but with radial boundaries determined by the prograde and retrograde photon orbit radii (equations \ref{eq:r_p} and \ref{eq:r_r}). The region outside of the retrograde photon orbit contributes almost $90\%$ of the total flux in the images. However, the region between the prograde and retrograde photon orbits still contributes significantly to the central ring-like emission. Negligible emission originates from within the photon orbit.

\vspace{0.5em}

Since the examples above focused on a single GRMHD snapshot, those results are not necessarily representative of the simulations as a whole. To further explore the behavior in Figures~\ref{fig:3_color_panel} and \ref{fig:cross_sect_MAD}, we calculate the average image contribution from spherical shells around the black hole. In the bottom panels of Figure~\ref{fig:Changing_R_high_MAD} we calculate the average invariant synchrotron emission $j_{\nu} / \nu^2$ within concentric spherical shells of width $0.5\,M$ as a function of the mean shell radius for several values of $R_{\mathrm{high}}$, $\mathcal{M}$, and $i$. Increasing $R_\mathrm{high}$ results in an overall decrease of $j_{\nu} / \nu^2$. This is because synchrotron emissivity increases with $T_e$, and $T_e$ decreases (or stays approximately constant when $\beta \lesssim 1$) when $R_{\mathrm{high}}$ increases. In contrast, the flux increases with $\mathcal{M}$ since synchrotron emission depends on both the electron number density and the magnetic field, both of which increase as $\mathcal{M}$ increases (see equation~\ref{eq:Synchrotron_emission}). The emissivity peaks close to the prograde photon orbit and decreases monotonically at larger radii (peaks between $r\approx 2.35\,M$ and $r\approx 2.85\,M$ at $i=17^{\circ}$, and $r\approx 1.85\,M$ and $r\approx 2.85 \,M$ for all $i$).

\vspace{0.5em}

To explore the effects of absorption, we also plot the specific intensity $I_{\nu}/\nu^3$ in the top panels of Figure~\ref{fig:Changing_R_high_MAD}. Since the behavior of $j_{\nu} / \nu^2$ vs.~radius is similar to the behavior of $I_{\nu}/\nu^3$ vs.~radius, the optical depth must be relatively low. However, the simulation with the highest value of $\mathcal{M}$ is a notable counterexample. In that simulation, the larger accretion rate implies a larger number density, which leads to a higher optical depth, attenuating emission at $r \gtrsim 30M$. Note, however, that we ignore the effect of matter outside each shell, i.e., absorption outside of a shell is not included in the calculation of the average specific intensity contributed by the shell.

\vspace{0.5em}

Figure~\ref{fig:Changing_R_high_SANE} is similar to Figure~\ref{fig:Changing_R_high_MAD} but for a SANE simulation. Both MAD and SANE simulations behave quite similarly, with the MAD curves being slightly more smooth than SANE (emissivity still peaks between $r\approx 1.85\,M$ and $r\approx 2.85 \,M$ for all $i$ for SANE). Absorption has an even smaller effect in SANE simulations as compared to MAD. For both SANE and MAD, $j_{\nu} / \nu^2$ peaks at slightly smaller radii compared to $I_{\nu}/\nu^3$, indicating that absorption is not negligible very close to the black hole. For both MAD and SANE, the slopes of both $j_{\nu} / \nu^2$ and $I_{\nu}/\nu^3$ decrease as $\mathcal{M}$ increases. This behavior suggests that the outer regions of the flow become brighter relative to the inner regions. This relative change is likely due to the fact that the magnetic field scales with $\mathcal{M}$ and synchrotron emissivity scales non-linearly with magnetic field (see equation~\ref{eq:Synchrotron_emission}), resulting in a proportionately higher increase in emission farther from the black hole.

\vspace{0.5em}

For all values of $R_{\mathrm{high}}$, $\mathcal{M}$, and $i$ we consider for both MAD and SANE and both $j_{\nu} / \nu^2$ and $I_{\nu}/\nu^3$, the majority of the emission in 1.3~mm images originates within the retrograde ISCO radius and peaks between $r\approx 1.35\,M$ and $r\approx 3.35\,M$. This is consistent with previous expectations and provides additional certainty that the 1.3~mm images of M87 and Sgr~A$^*$ probe the region very close to the black hole, as has previously been argued (see, e.g., \citealt{M87_paper_5,Patoka}).

\vspace{0.8em}

The exploration we performed in this section only required running six ray-tracing simulations, one for each inclination angle shown in Figures~\ref{fig:Changing_R_high_MAD} and \ref{fig:Changing_R_high_SANE}. These six ray tracing simulations generated over 1,200 images, which were used for the averages shown in the figures (100 snapshots in time and 12 parameter combinations). Once we compute the trajectory information, the radiative transfer calculations are vectorized and are thus very efficient.

\section{Summary}\label{sec:conclusion}
We present {\tt Mahakala}, a new, {\tt Python} based, modular ray-tracing and radiative transfer code for arbitrary space-times. {\tt Mahakala} uses Google's new machine learning framework, {\tt JAX}, which efficiently parallelizes computation on CPUs, GPUs, and TPUs. {\tt JAX} performs accelerated automatic differentiation, allowing the user to work in arbitrary space-times without the need to manually calculate Christoffel symbols. {\tt Mahakala} has been developed to simulate the mm-wavelength radiation of low-luminosity accreting supermassive black holes and calculates synchrotron emission and absorption. The code uses Cartesian KS coordinates to avoid the numerical issues near the poles that arise in spherical coordinate systems, and it can integrate photon trajectories with respect to either coordinate time or affine parameter according to the user's choice. {\tt Mahakala} natively supports the new GPU-accelerated {\tt AthenaK} GRMHD code, which also uses Cartesian KS coordinates. 

We verify both the radiative transfer and geodesic integration components of {\tt Mahakala} with tests from the literature (see Section~\ref{sec:Tests}). We show that the errors in the deflection angle of null trajectories near a Kerr black hole are in the range of other radiative transfer codes explored in \citet{Gold}.  We perform convergence tests with spherical photon orbits and show that {\tt Mahakala} converges as expected for a fourth order scheme (see also \citealt{GRay2}). Finally, we test the radiative transfer component of {\tt Mahakala} with analytic accretion model tests from \citet{Gold}. The image morphology and total flux are consistent with the results of several radiative ray-tracing algorithms that are compared in \citet{Gold}.

One of the main design aims of {\tt Mahakala} is flexibility and ease of use. Since {\tt Mahakala} can easily be run in a {\tt Python jupyter notebook}, we hope that it will lower the barrier to entry for radiative ray-tracing simulations. The modular nature of {\tt Mahakala} allows us to explore in detail how different regions of the 3-D GRMHD flow volume contribute to image features. We demonstrate this capability with example snapshots from MAD and SANE GRMHD simulations with $a=0.9375\,M$ generated by the {\tt AthenaK} code. We show in Section~\ref{sec:applications} that the majority of emission at 1.3~mm comes from within $\sim10\,M$ of the black hole with  emissivity peaks between $r\approx 1.85\,M$ and $r\approx 2.85 \,M$ for all models. This result is robust to moderate changes in the mass accretion rate, the $R_{\mathrm{high}}$ parameter that sets the electron temperature prescription, and the observer's inclination angle with respect to the black hole spin axis $i$. These results provide further evidence that the EHT images of M87 and Sgr~A$^*$ probe the regions close to their respective black holes and that most of the emission has been gravitationally lensed by the black holes. 

In addition to the dependence on the emission radius, we also explore how conical shells at different $\theta$ contribute to the image. For our example MAD simulation, we find that the disk contributes most of the emission followed by the forward and counter jets, respectively. The forward jet contributes some emission to the center of the ring feature and also to the region above the ring. In future work, we will extend {\tt Mahakala} to include polarization and relax the fast light assumption to allow the fluid to evolve as light propagates through the domain.


\vspace{.2in}

The authors thank the anonymous reviewer for helpful suggestions. L.\;M.\ gratefully acknowledges support from a NASA Hubble Fellowship Program, Einstein Fellowship under award number HST-HF2-51539.001-A, an NSF Astronomy and Astrophysics Postdoctoral Fellowship under award no. AST-1903847, and AST-2407810.
C.\;C.~acknowledges support from NSF Partnerships for International Research and Education (PIRE) and Mid-Scale Innovations Program (MSIP) grants OISE-1743747 and AST-2034306. 
G.\;N.\;W.~was supported by the Taplin Fellowship. G.\;H. recognizes support from a Visiting Scholarship at CIERA/Northwestern University.
All ray tracing and radiative transfer calculations were performed with the {\tt ElGato} (funded by NSF award 1228509) and {\tt Ocelote} clusters at the University of Arizona. This work has been assigned a document release number LA-UR-23-23447.



\bibliography{main}

\end{document}